\documentclass{aastex}
\newcommand{\etal}{\it et al.}

\shorttitle{PHOTOMETRIC SURVEY OF  OPEN CLUSTERS}
\shortauthors{Ann \it{et al.}} 

\begin{document}
 
\title{BOAO PHOTOMETRIC SURVEY OF GALACTIC OPEN CLUSTERS. II. 
Physical Parameters of 12 Open Clusters}

\author{H. B. Ann and S. H. Lee}
\affil{Department of Earth Science, Pusan National University, Pusan 609-735, Korea (hbann@hyowon.pusan.ac.kr, shlee@jupiter.es.pusan.ac.kr)}

\author{H. Sung and M. G. Lee}
\affil{Astronomy Program, SEES, Seoul National University, Seoul 151-742, Korea (sungh@astro.snu.ac.kr, mglee@astrog.snu.ac.kr)}
\author{S. -L. Kim, M. Y. Chun, Y. -B. Jeon, B. -G. Park, and I. -S. Yuk}
\affil{Korea Astronomy Observatory, Taejeon, 305-348, Korea (slkim@kao.re.kr,
mychun@kao.re.kr, ybjeon@boao.re.kr, bgpark@boao.re.kr, yukis@kao.re.kr)}

\begin{abstract}
We have initiated a long-term project, the BOAO photometric survey of 
open clusters,
to enlarge our understanding of galactic structure using $UBVI$ CCD 
photometry of open clusters which have been little studied before.
This is the second paper of the project in which we present the photometry
of 12 open clusters. We have determined the cluster parameters by
fitting the Padova isochrones to the color-magnitude diagrams of the clusters.
All the clusters except for Be 0 and NGC 1348 are found to be 
intermediate-age to old (0.2 -- 4.0 Gyrs) open clusters with a mean
metallicity of [Fe/H]$ \approx 0.0$.

\end{abstract}

\keywords{Open clusters and associations : general -- Open clusters and associations : 
individual (Be 104, Be 60, King 15, NGC 381, Be 64, King6, NGC 1348,
Be 12, Be 70, Be 23, NGC 2259 and NGC 2304) -- photometry : distance, age}
 
\section{INTRODUCTION}
 
Open clusters are an ideal tool for tackling many astrophysical problems such 
as star formation processes, stellar evolution and dynamical evolution
of stellar systems.
To date there are about 1200 known galactic open clusters in the
Catalog of Open Cluster Data (Lyng\aa~1987; hereafter COCD).
Owing to the high quantum efficiency, good linearity and large
dynamic range of modern CCDs, 
accurate color-magnitude diagrams (CMD) of a large 
number of open clusters have been obtained to derive their basic physical
parameters (distance, interstellar reddening, age and 
metallicity). Current understanding of the galactic disk has greatly
benefited from several recent systematic CCD observations of young and old
open clusters \citep{phe93,jan94}.
 
However, more than half of the known galactic open clusters are as yet
virtually unstudied. All that we know about the majority of the open clusters
are their positions and angular diameters, which are often
inaccurate in many cases. To date only one third of the known open
clusters have been studied in detail to derive their basic parameters.
Moreover, more than half of the well studied clusters
belong to Trumpler richness class $m$ and $r$, while most of the unstudied
clusters are classified as $p$. Thus, present knowledge of open clusters
is considered to be heavily biased by the properties of rich open clusters.

To understand the nature of open clusters and the structure of the
galactic disk as a whole, we have started a comprehensive photometric
survey of open clusters which have been little studied before,
using the 1.8m telescope at
Bohyunsan Optical Astronomy Observatory (BOAO) in Korea. 
The primary goal of the present survey is to obtain accurate CMDs of the 
unstudied clusters using $UBVI$ CCD photometry
and to derive cluster parameters such as distance and age. 
We selected 343 clusters by the declination
criterion of  $\delta > -20^{\circ}$.  They comprise about half of
the clusters with unknown distance and age in COCD.
A more detailed description of the present survey was given by 
\citet{ann99}.
  
This is the second paper of our survey project. Here, we 
present the results of the photometric survey of 12 open clusters observed
during the observing runs in 1998 and 1999.
In \S2, we describe the observations and data reduction. The  
physical parameters of the observed clusters are presented in \S3 and
a brief summary and discussion are given in the last section.

\section{OBSERVATIONS}

We observed 12 candidate clusters of the BOAO photometric survey 
in December 1998. The basic parameters of the observed
clusters \citep{lyn87} are given in Table 1. The 
photometric observations were made using the SITe $2048 \times 2048$ CCD 
camera with Johnson-Cousins $UBVI$ filters attached to the F/8 cassegrain 
focus of the BOAO 1.8m telescope. 
The pixel size of the CCD is 24 $\mu\rm{m}$ 
corresponding to 0.$^{\prime \prime}$34 on the sky. 
The gain and readout noise of the CCD are 1.8 electrons/ADU and 7 electrons,
respectively. 
We have obtained images of the central $11.^{\prime}6 \times 11.^{\prime}6$
regions of each cluster with a pair of long and short exposures
for each filter. We observed several $UBVI$ standard stars in Landolt (1992)'s
catalog for the calibration of our photometry during the observing run.  
However, owing to 
non-photometric conditions and difficulty in transforming the $U-B$ colors
to the standard system, we re-observed the clusters in a short exposure with 
the SAAO version of Landolt standard stars in the equator \citep{men91}
and blue and red standard stars \citep{kil98} on a photometric night
during the observing run in October 1999.
The observed standard stars (or regions) are SA 112, SA 114, SA 93, SA 99,
BD -11 162, GL 876 and GL 273. Table 2 lists the observational log.

Grayscale maps of the $V$-band images of the clusters 
taken with long exposures are displayed in Fig.~1. 
In most cases, the cluster field is a small part of the observed frame.
Solid circles in Fig.~1 represent the angular diameters of the clusters given
in the COCD whose
centers were determined by visual inspection
of the images. We obtained photometric diagrams of the
clusters for the stars located within the circles.

We have followed standard CCD reduction techniques 
using IRAF/{\small CCDRED}
for the preprocessing of the images obtained.  
This includes the subtraction of bias levels with
overscan correction, trimming and flat-fielding.  
We used IRAF/DAOPHOT 
to obtain instrumental magnitudes of the stars.
We applied aperture corrections to the instrumental 
magnitudes of cluster stars obtained from Point Spread Function (PSF) fitting,
using the instrumental magnitudes obtained with an aperture radius 
of $7^{\prime \prime}$.
 
The transformation of the instrumental system to the standard system was 
made using the following equations
(see the similar transformation for the $40^{\prime \prime}$ 
telescope at Siding Spring
Observatory: \citep{sun00}):
$$V = v - k_V X - 0.046 (B-V) + \zeta_V, $$
or
$$V = v - k_V X - 0.046 (V-I) + \zeta_V'~~~~~for~ V-I < 1.5, $$ 
$$V = v - k_V X + \zeta_V''~~~~~for~ V-I \geq 1.5, $$ 
$$I = i - k_I X - 0.017 (V-I) + \zeta_I~~~~~for~ V-I < 0.6, $$
$$I = i - k_I X + 0.083 (V-I) + \zeta_I'~~~~~for~ V-I \geq 0.6, $$
$$B = b - k_B X + 0.112 (B-V) + \zeta_B,$$
$$U = u - k_U X + f(B-V) + \zeta_U $$
where uppercase characters, lowercase characters, $k$, X and $\zeta$ represent
standard magnitudes, instrumental magnitudes, extinction coefficients, airmass,
and zero points, respectively. 
In the transformation of the $U$ magnitude, a non-linear function, $f(B-V)$,
is involved.
The function $f(B-V)$ allows the correction of redder $(U-B)$ colors, probably
due to a red-leak in the $U$ filter \citep{bes01}. 
It increases linearly up to $(B-V) \approx 0.8$,
but shows a non-linear increase for redder stars.
This correction
dominated the other factors, and we did not take into account the ($U-B$)
term in the $U$ transformation. The extinction coefficients derived from
standard star photometry are $k_U$ = 0.485 $\pm$ 0.024, $k_B$ = 0.232$\pm$
0.035, $k_V$ = 0.142 $\pm$ 0.023 and $k_I$ = 0.058 $\pm$ 0.027. The final
transformation errors were between 0.021 mag in $V$ and 0.038 mag in $U$.
The instrumental magnitudes obtained from the observations in non-photometric 
conditions during the December run were transformed to the standard system
using the observations of the October run.  The photometric data will be
presented in the electronic AJ.

We calculate the external observational errors of the photometry of the 
cluster stars using the pair of exposures
in the 1998 December run and the short exposures in the 1999 October run.
Fig. 2 shows the resulting observational errors as a function of $V$ magnitude.
The error bars in Fig. 2 are the standard deviations of the observational
errors for the stars in each magnitude bin. As shown in Fig. 2, the 
observational errors for stars fainter than $V \approx 14$
increase with increasing magnitude, while the observational errors for
stars brighter than $V \approx 14$ are nearly constant with relatively
large uncertainties due to the small number of stars in the magnitude bins.

\section{RESULTS}

We have determined the basic physical parameters of the 12 open clusters
by analyzing the color-color diagrams and CMDs of the clusters.
Photometric diagrams of the clusters are derived from the photometry
of stars 
inside the angular diameters \citep{lyn87} plotted in 
Fig. 1. To achieve improved accuracy in the derived parameters, 
we used stars with DAOPHOT errors less than 0.1 mag.
In addition, due to large errors in $U$, the $(U-B)$ vs $(B-V)$ diagrams
were constructed using stars brighter than $V \approx 17$. 
In general, as shown in Fig. 2, the total observational errors
are quite large ($\sigma_{v} \approx 0.03$) even for the 
brightest stars. 
However, the present photometry is good enough to derive the 
cluster parameters using isochrone fittings.

We have estimated the reddening of the clusters from $(U-B)$ vs $(B-V)$ 
diagrams,
by comparing theoretical ZAMS of the Padova isochrones \citep{ber94} with the
observed distributions of cluster stars. To do this, we assumed 
a slope of $E(B-V)/E(U-B)$ = 0.72 \citep{jon53},
a total to selective extinction ratio ${R_V}$=3.0 and
$E(V-I)=1.25 E(B-V)$ \citep{dea78}. The advantage of using theoretical ZAMS
is the ability to take into account the metallicity effect on the 
stellar distribution in the color-color diagrams. Because the
metallicity of clusters is not known beforehand, we assumed the solar abundance
as the cluster abundance unless the stellar distributions in the color-color
diagrams and CMDs can not be described by the solar abundance isochrones. 

We used the Padova isochrones \citep{ber94} in the isochrone
fittings. We put the greatest weight on the
stars in the turnoff regions and giant branches, if present, since the
lower main sequences are more heavily contaminated by field stars.
We attempted to fit the lower envelope of the main sequence,
considering the photometric errors. 
The best fit solutions were obtained by visual inspections of the
CMDs with the trial fittings.  
We present the results of the best fits in Fig. 3 and the resulting 
cluster parameters are given in Table 3. 

Since we put the greatest weight on the stars in the regions which are 
supposed to be the giant branches of some clusters, the memberships of
these stars are crucial to the derived cluster parameters. To see whether
these stars are cluster giants or not, we plotted cleaned CMDs in Fig. 4. 
To obtain the cleaned CMDs, 
we statistically removed field stars from the cluster CMDs by finding
stars nearest to the field stars with a photometric box of 
$\pm 0.25$ in $V$ and $\pm 0.1$ in $(B-V)$.
Because we did not observe field region of each cluster, we constructed
the CMDs of field stars by selecting stars randomly
from the regions outside the radii that are two times larger than the
cluster radii in the same CCD images, except for 
Be 70 for which field stars are selected from the region outside 1.5 times 
larger than the cluster radius due to the larger angular size of the cluster.
As shown in Fig. 4, giant stars are left over in the cleaned CMDs.
This result strongly suggests that many stars in the red giants locus
could be the members of the cluster. 
In the following, we 
briefly describe the properties of the individual clusters.

\subsection{Berkeley 104}

Be 104 is located to the south of a very bright star, HD225094.
As indicated by its Trumpler class of $II1p$, Be 104 is one of the most 
poorly defined clusters. Due to the paucity of members and the heavy 
contamination
by field stars, the CMDs of Be 104 show a very broad main sequence which
is not much different from that of the field star CMDs near the cluster.
Thanks to the presence of a few giants which are thought to be
cluster members as they appear in the cleaned CMD in Fig.4, 
we could derive
the cluster parameters by fitting the isochrones to the stellar
distributions in the CMDs. 
The resulting cluster parameters are
$(m-M)_{0}=13.2 \pm 0.2$, $E(B-V)=0.45 \pm 0.05$, log(t[yrs]) = $8.9 \pm0.1$, 
and [Fe/H] $= 0.07$ dex. Most of the stars brighter than the turnoff point
of the main-sequence stars are considered to be field stars. However, there
seem also to be a few blue straggler candidates according to their positions 
on the CMDs.

\subsection{Berkeley 60}

Be 60 is also a poorly defined cluster with a Trumpler class of $III1p$.
As shown in Fig. 1, there are several bright stars in the field of Be 60
most of which seem to be foreground stars. However, the CMDs of the stars
in the central region of Be 60 show 
a well defined main sequence which permits a reasonably good isochrone fitting.
The resulting best fit isochrone 
gives the cluster parameters as $(m-M)_{0}=13.2 \pm 0.2$,
$E(B-V) = 0.86 \pm 0.05$, log(t[yrs]) = $8.2 \pm 0.2$ and [Fe/H] $= 0.07$ dex.
The two brightest red stars near $(B-V) =1.6$ are not plotted in 
the $V$ vs $(V-I)$ CMD due to saturation in the $I$ image.

\subsection{King 15}

As shown in Fig. 1, King~15 shows the typical properties of $IV2p$ clusters,
which are characterized by a small number of member stars and no central
concentration. However, as can be seen in Fig. 3,
the stellar distributions in the CMDs of the cluster
are well represented by the isochrone of log(t[yrs]) = $8.4 \pm 0.1$ and 
[Fe/H] = $0.07$ dex with a distance modulus and reddening of
$(m-M)_{0}=12.5 \pm 0.4$ and  $E(B-V) = 0.70 \pm 0.12$, respectively.
\citet{phe94} derived a slightly smaller distance modulus,
$(m-M)_{0} = 12.29$ and smaller reddening, $E(B-V) = 0.46$ for King 15. 
Since they had no $U$-band photometry, their reddening estimate was obtained
by fitting the \citet{sch82}'s ZAMS to the stellar distribution in the 
$V$ vs $(B-V)$ CMD. 
They estimated the age of King 15 to be $\sim 3$ Gyr from the color of the
earliest spectral type star on the main-sequence ($(B-V)_{o}=0.19$).
The reason for the large discrepancy between the age derived from the isochrone
fittings and that found by \citet{phe94} is due to the fact that they considered
stars brighter than $V \approx 15$ to be field stars, while we consider
them as cluster members according to their positions in the CMDs as well
as spatial position in the cluster field. 

As shown in Fig. 4,
the presence of stars brighter than $V \approx 15$ in the cleaned CMD 
support the assumption that $V \sim 14$ are cluster members.
However, it is quite natural for them 
to believe the stars brighter than $V \approx 15$ to be field
stars since their CMD is heavily populated by field stars
due to the larger field compared to ours.
They constructed the CMD from the stars in a field of 
$11.^{\prime}6 \times 11.^{\prime}6$, which is nearly the same field as 
that of the total CCD image for King 15 in Fig. 1. 
If stars brighter than $V \approx 15$ are member stars of King 15, 
there is a large gap in the upper main sequence of the cluster.

\subsection{NGC 381}
NGC 381 is an intermediate-age open cluster with a Trumpler class of $III1m$.
There seem to be no giant stars in the cluster probably due to its small number
member stars. The main sequence of the cluster  is well defined in the CMDs
although there are a large number of field stars in the direction of NGC 381.
Owing to the foreground and background field stars, the color-color diagram
of NGC 381 is quite complicated. However, it is not difficult to determine 
the cluster reddening by fitting the theoretical ZAMS to the distribution of
the main-sequence stars in the color-color diagram because they are well
isolated from the field stars. As shown Fig.3,
the isochrone of log(t[yrs]) = $8.5 \pm 0.1$ and [Fe/H] $= 0.07$ dex fits well
the stellar distribution in the CMDs of NGC 381 with a distance modulus
and reddening of $(m-M)_{0}=10.3 \pm 0.3$ and $E(B-V) = 0.40 \pm 0.10$,
respectively. The present estimates of the distance modulus and reddening 
of the cluster are in good agreement with those determined by \citet{phe94} but
their age estimate of log(t[yrs]) = 9.04, obtained using the isochrone 
of \citet{mae91}, is about a factor of three larger than ours.

\subsection{Berkeley 64}
Be 64 shows  the typical morphology of a Trumpler class $II1m$ cluster
which is characterized by a slight concentration of member stars of 
similar brightness with medium richness. Because the color-color diagram 
does not show any clear cluster sequences due to the small number of bright
stars with small photometric errors in $U$, we adopt 
$E(B-V) = 1.05$ \citep{pan97} as the cluster reddening. 
The CMDs of Be 64 show a quite broad
main sequence due to contamination by field stars 
However, the presence of giant stars, most of which are thought to
be cluster members from the cleaned CMD in Fig. 4,
makes it possible to fit the isochrones to the stellar
distributions in the CMDs unambiguously. 
As shown in Fig. 3, the adopted isochrone of log(t[yrs]) = $9.0 \pm 0.1$ and
[Fe/H] = $-0.61$ dex with $(m-M)_{0}=13.0 \pm 0.3$ and $E(B-V) = 1.05 \pm 0.15$
fits the stellar distribution in the CMDs quite well. Our estimates of
the distance modulus and age are in good agreement with those of \citet{pan97}.

\subsection{King 6}

King 6 is a nearby cluster with a Trumpler class of $II2m$.
As shown in Fig. 1, the angular diameter of the cluster is comparable 
to the size of the observed field because of its proximity.
The CMDs of King~6 show a well defined main sequence with a relatively
small number of field stars, especially in the upper main sequence.
However, the lower main sequence, fainter than
$V \sim 18$, seems to be contaminated by field stars. Although there
seems to be a clear sequence in Fig. 3, 
the stellar distributions in the color-color diagram of King 6 can not 
be fitted by the theoretical ZAMS with a single reddening value. 
The stars fainter than $V \sim 16$ seem to have a smaller reddening of
$E(B-V) \approx 0.4$, while the stars brighter than $V \sim 13$ appear
to have a larger reddening of $E(B-V) \approx 0.6$. Thus, we adopted a
mean value of $E(B-V)=0.5 \pm 0.10$ as the reddening of King 6. This is
the reason why the stellar distributions in the color-color diagram show a
relatively poor match with the theoretical ZAMS which is shifted by a reddening
value of $E(B-V)=0.5$. However, as shown in Fig. 3, the isochrone of
log(t[yrs]) = $8.4 \pm 0.1$ and [Fe/H] $= 0.46$ dex, 
shifted by a distance modulus and reddening
of $(m-M)_{0}=9.7 \pm 0.4 $ and $E(B-V) = 0.5 \pm 0.10$, respectively,
gives a pretty good match
with the stellar distributions in the CMDs of the cluster.
There is a slight indication of the presence of the
binary sequence above the ZAMS.

\subsection{NGC 1348}
NGC 1348 is the youngest among the observed clusters and is 
relatively well recognized due to the absence of field stars comparable
in brightness to the bright members of the cluster, although the cluster
stars do not show any obvious central concentration. This is the reason why NGC 1348 is
classified as Trumpler class $III2m$.
As shown in Fig. 3, the color-color diagram of NGC 1348 shows a well defined
sequence which is well matched with the theoretical ZAMS of the solar abundance
 shifted by 
$E(B-V)=1.02$. The isochrone fittings to the stellar distributions in the
CMDs give the cluster parameters of log(t[yrs]) = $8.1 \pm 0.1$,  
$(m-M)_{0}=11.3 \pm 0.3$, $E(B-V)=1.02 \pm 0.05$ and [Fe/H] $=0.07$ dex, 
respectively.
The brightest star near $B-V$=2.4
is saturated in the $I$ image and is omitted in the $V$ vs $(V-I)$ diagram.

\subsection{Berkeley 12}

Be 12 is an old cluster with a Trumpler class of $II1m$. Although Be 12
is located in the direction of a rich star field, it can be well
distinguished from
field stars due to the concentration of the member stars. The CMDs
of the cluster show a broad main sequence caused by heavy contamination by 
field stars. The color-color diagram of Be 12 also shows no clear sequence
due to the lack of a sufficient number of cluster main-sequence stars that
are bright enough to have small photometric errors in $U$ as well as the 
contamination by field stars. The presence of blue straggler stars which
are located in the bright extension of the main sequence of Be 12 may 
contribute to the large scatter in the color-color diagram. 
However, as shown in Fig. 3,
the isochrone of log(t[yrs]) $= 9.6 \pm 0.1$ and [Fe/H] $=0.07$ dex fits 
the main sequence together with the giant branch quite well
with a distance modulus and reddening of $(m-M)_{0}=12.5 \pm 0.5$ and
$E(B-V)=0.7 \pm 0.15$, respectively. The old age of the cluster supports
the existence of blue straggler stars based on the presence of 
stars brighter and bluer
than the turnoff point of the cluster main sequence 
because such stars are more frequently found in old open
clusters \citep{ahu95}

\subsection{Berkeley 70}

Be 70 is located in a direction toward which a large number of 
foreground stars are located. The CMDs of the cluster show  a broad
main sequence due to heavy contamination by field stars.  
The color-color diagram appears to be complicated due to the field star
contamination, but we could determine the reddening of the cluster by fitting
the theoretical ZAMS to the distributions of upper main-sequence stars and
populous giant stars which constitute a well defined cluster sequence in
the CNDs. The broad sequence around $(B-V) \approx 0.7$ in
the color-color diagram is due to the field star population, whose mean 
reddening is much smaller than that of the cluster. The heavy contamination
of field stars is apparent from the cleaned CMD of Be 70 in Fig. 4.
However, some of the bright blue stars above the turnoff point of 
the main sequence of Be 70 may be blue straggler stars from their presence
in the cleaned CMD of the cluster.
Owing to the well developed giant branch and the highly
populated main sequences, it is not difficult to fit the isochrones to 
the stellar distributions in the CMDs. The resulting cluster parameters are
log(t[yrs]) $= 9.6 \pm 0.1$, [Fe/H] $=-0.32$ dex, $(m-M)_{0}=13.1 \pm 0.3$ 
and $E(B-V)=0.48 \pm 0.10$. 

\subsection{Berkeley 23}

Be 23 is located near the direction of the anti-galactic center direction. 
As indicated by its Trumpler class of $III1m$,
it shows no strong central concentration but can be identified by its
relatively dense population compared to that of the field stars.
Most of the brightest stars above the turnoff point are considered to
be foreground stars because the CMD of a nearby field region shows similar
stellar distributions. 
However, there is also a possibility that some of the stars that are
brighter and bluer than the turnoff of the main sequence are blue
straggler stars.  We derived the cluster 
parameters by fitting the isochrone of log(t[yrs]) $= 8.9 \pm 0.1$
and [Fe/H] $=0.07$ to the stellar distribution, especially near the 
turnoff, with a distance modulus and reddening of
$(m-M)_{0}=14.2 \pm 0.3$ and $E(B-V)=0.40 \pm 0.05$, respectively. 
The theoretical ZAMS of the solar abundance shifted by the adopted reddening
of $E(B-V)=0.40$ describes well the
distribution of upper main-sequence stars in the color-color diagram.
Most of the forground stars seems to be affected by negligible reddening.

\subsection{NGC 2259}

NGC 2259 is a well detached system of Trumpler class $II1p$.
Most of the bright stars in the direction of NGC 2259 are cluster
main-sequence stars near the main-sequence turnoff. There seems to be
little contamination by field stars in the upper main sequence but the 
lower main sequence is somewhat contaminated by field stars.
Owing to the proximity
of the cluster, the main-sequence stars bluer than $(B-V) \approx 1.0$
are bright enough to produce a clear sequence in the color-color diagram.
Thus, we derived the cluster reddening by fitting the theoretical ZAMS
to the distribution
of bright main-sequence stars in the color-color diagram. 
We applied the isochrone fittings to the stellar distributions in the CMDs
of NGC 2259 with the reddening derived from the ZAMS fitting. As shown in 
Fig. 3, the isochrone of log(t[yrs]) $= 8.5 \pm 0.1$ and [Fe/H] $=0.07$ dex 
with a distance modulus and reddening of $(m-M)_{0}=12.6 \pm 0.3$ 
and $E(B-V)=0.59 \pm 0.05$, respectively,
matches well with the stellar distributions in the CMDs of NGC 2259.

\subsection{NGC 2304}

NGC 2304 is located in the direction of the anti-galactic center, slightly
above the galactic plane.
It is well distinguished from the field stars.
The color-color diagram shows a well defined cluster sequence which is
nicely matched with the theoretical ZAMS of [Fe/H] $=-0.32$ 
shifted by $E(B-V)=0.10$.
However, as shown in Fig. 3, the main sequence, especially the lower
main sequence, is somewhat contaminated by field stars. 
The stellar distributions in the CMDs of NGC 2304 show the typical 
characteristics of intermediate-age open clusters. The stars near the
giant branch seem to be cluster members since all these stars are present
in the cleaned CMD of NGC 2304 in Fig. 4. The isochrone
of log(t[yrs]) $= 8.9 \pm 0.1$ and [Fe/H] $=-0.32$ dex 
with $(m-M)_{0}=13.0 \pm 0.1$ and $E(B-V)= 0.10 \pm 0.02$
 fits well the stellar distribution in the CMDs of NGC 2304.
There seems to be a binary sequence above the ZAMS.

\section{DISCUSSION AND SUMMARY} 

We have determined the distances, reddenings, ages and metallicities
of 12 target clusters by isochrone fittings complemented by ZAMS fittings
for the reddening estimates. To our best knowledge, no known cluster parameters
were previously derived for any of the clusters except
for King 15, NGC 381, and Be 64.
Comparison of the cluster parameters derived from the present photometry 
with those found by other researchers
for King 15, NGC 381 and Be 64 \citep{phe94, pan97} shows that
our estimates are in good agreement with independent studies.

The mean interstellar absorption derived from the reddening of the 12
open clusters is $0.72 \pm 0.51$ mag$\cdot$kpc$^{-1}$, which is slightly
smaller than the mean interstellar absorption of the galactic disk,
$\sim 1.0$ mag$\cdot$kpc$^{-1}$, derived from COCD \citep{lyn87}.
The smaller mean interstellar absorption reflects the general characteristics
of the interstellar reddening toward the anti-galactic center, and the large
standard deviation is due to the inhomogeneous distribution of the interstellar
clouds \citep{luc78}. Be 23 and NGC 2304, which are located in a
direction very close to the anti-galactic center, show extremely small 
interstellar absorptions of 0.17 mag$\cdot$kpc$^{-1}$ and 0.08 
mag$\cdot$kpc$^{-1}$, respectively, while King 6 and NGC 1348, which are
located in the direction of the Cassiopeia-Perseus region \citep{deu76}, 
show large interstellar absorptions of 1.72 mag$\cdot$kpc$^{-1}$ and
1.68 mag$\cdot$kpc$^{-1}$, respectively. The large interstellar absorption
in this direction might be due to the dense interstellar cloud close to 
the Sun \citep{deu76}.

The apparent gap in the main sequence of King 15 seems to be a real
feature because the stars brighter than $V \approx 15$ are thought
to be cluster members. Thus, the age of King 15 is $\sim 250$ Myr, which
is much smaller than that suggested by \citet{phe94}.
However, the gap near $V \sim 14$ is not a 
physical gap related to stellar evolution but is 
caused by random statistical fluctuations due to the small number of 
member stars. King 15 is too young for the cluster to have a
main-sequence gap similar to those of NGC 2420 \citep{ant90,lee99} and
NGC 2243 \citep{bon90}, which are thought to be related to the 
rapid hydrogen exhausting phase of the convective core \citep{dem94}. 
The gap in King 15 is similar to those of Be 7, NGC 637, and 
Be 62 \citep{phe93}, all of which are thought to be caused by 
statistical fluctuations in the high mass part of the IMFs \citep{sca86}.

We have found blue straggler candidates in the four open clusters, 
Be 104, Be 12, Be 70, and Be 23, which have ages greater than 800 Myr.
There is a tendency for older clusters to have more blue straggler stars,
similar to that found by \citet{ahu95}.

Our results for the 12 clusters presented in this study
along with the results of \citet{ann99}
are expected to be very useful for the investigation of the open cluster system
in our Galaxy. Our survey project is still in progress and we plan to 
present a study of more open clusters in the near future.

The authors would like to thank to KAO for the allocation of telescope times.
The work by H.B.Ann was supported in part by PNU grant 1999-2003.

\newpage
\figcaption{Annfig1.gif}{A mosaic map of grayscale $V$-band images of 12 open 
clusters. The circles represent
the angular diameters of the clusters given in the COCD \citep{lyn87}.
The size of each box is $11.^{\prime}6 \times 11.^{\prime}6$. North is up
and east is to the left.
}

\figcaption{Annfig2.gif}{Observational errors in the present photometry.
The error bars represent the standard deviations of observational errors
for stars in each magnitude bin of 0.5 mag.
The large error bars for
the brighter stars are caused by the small number of stars in the
magnitude bin. 
}

\figcaption{Annfig3.gif}{Photometric diagrams for the 12 open clusters.
The dashed and solid lines in the color-color diagrams represent 
the theoretical ZAMS of the Padova isochrones\citep{ber94}. The solid lines in
CMDs are the best fit isochrones.
}

\figcaption{Annfig4.gif}{The cleaned CMDs of the 6 open clusters 
that are supposed
to have giant members. The best fit isochrones
shifted according to the derived reddening and distance moduli
are plotted.
}
\newpage

\pagestyle{empty} 
\begin{table*}
\begin{center}
\tablenum{1}
\caption{Basic parameters of the 12 open clusters. \label{tbl-1}} 
\begin{tabular}{ccccccc}
\tableline\tableline
 Cluster   & R.A.(2000.) &  DEC.(2000.) &     l    &    b    & Trumpler class & Angular Diameter \\
\tableline
 Be  104  & ~0 ~3 28.1  &  63 35 42.4  &  117.63  &  ~1.22  &    ~II 1 p     & $~3.^{\prime}0$ \\
 Be   60  & ~0 17 41.4  &  60 57 40.3  &  118.85  &  -1.64  &    III 1 p     & $~3.^{\prime}0$ \\
 King   15  & ~0 32 55.8  &  61 51 33.8  &  120.75  &  -0.95  &    ~IV 2 p     & $~3.^{\prime}0$ \\
 NGC  381  & ~1 ~8 19.9  &  61 35 ~2.1  &  124.94  &  -1.22  &    III 1 m     & $~6.^{\prime}0$ \\
 Be   64  & ~2 21 ~2.8  &  65 53 48.3  &  131.92  &  ~4.60  &    ~II 1 m     & $~2.^{\prime}0$ \\
 King    6  & ~3 28 ~3.5  &  56 27 30.1  &  143.35  &  -0.07  &    ~II 2 m     & $10.^{\prime}0$ \\
 NGC 1348  & ~3 33 51.7  &  51 26 ~9.5  &  146.94  &  -3.71  &    III 2 m     & $~5.^{\prime}0$ \\
 Be   12  & ~4 44 37.4  &  42 40 38.3  &  161.68  &  -1.99  &    ~II 1 m     & $~4.^{\prime}0$ \\
 Be   70  & ~5 25 44.6  &  41 53 44.6  &  166.90  &  ~3.58  &    III 1 m     & $~6.^{\prime}0$ \\
 Be   23  & ~6 33 28.5  &  20 32 47.0  &  192.61  &  ~5.44  &    III 1 m     & $~4.^{\prime}0$ \\
 NGC 2259  & ~6 38 34.4  &  10 53 24.1  &  201.79  &  ~2.12  &    ~II 1 p     & $~3.^{\prime}0$ \\
 NGC 2304  & ~6 55 ~0.9  &  18 ~1 14.1  &  197.17  &  ~8.87  &    ~II 1 m     & $~3.^{\prime}0$ \\
\tableline
\end{tabular}
\end{center}
\end{table*}

\newpage
\pagestyle{empty}
\begin{table*}
\begin{center}
\tablenum{2}
\caption{Observational log of the 12 open clusters. \label{tbl-2}}
\begin{tabular}{cccccc}
\tableline\tableline
Cluster & Filter  &  T$_{\rm exp}$        & Cluster & Filter  &  T$_{\rm exp}$ \\
\tableline
 Be 104    & U  & $200 s \times 4$        & Be 60     & U  & $200 s \times 3$       \\
           & B  & $200 s \times 3$, 50 s  &           & B  & $200 s \times 4$       \\
           & V  & $100 s \times 3$, 10 s  &           & V  & $100 s \times 3$, 50 s \\
           & I  & $50 s \times 3$, 5 s    &           & I  & $50 s \times 3$        \\
\tableline
 King 15   & U  & $200 s \times 3$        & NGC 381   & U  &  900 s, 90 s           \\
           & B  & $200 s \times 3$, 20 s  &           & B  & 400 s, 20 s            \\
           & V  & $100 s \times 3$, 10 s  &           & V  & 130 s, 10 s            \\
           & I  & $50 s \times 3$, 5 s    &           & I  & 50 s, 5 s              \\
\tableline
 Be 64     & U  &  900 s                  & King 6    & U  & $200 s \times 3$       \\
           & B  & 900 s                   &           & B  & $200 s \times 3$, 20 s \\
           & V  & 400 s                   &           & V  & $100 s \times 3$, 10 s \\
           & I  & 150 s 30 s              &           & I  & $50 s \times 3$, 5 s   \\
\tableline
 NGC 1348  & U  & $200 s \times 4$        & Be 12     & U  &  900 s                 \\
           & B  & $200 s \times 3$, 20 s  &           & B  & 900 s                  \\
           & V  & $100 s \times 3$, 10 s  &           & V  & 400 s                  \\
           & I  & $50 s \times 3$, 5 s    &           & I  & 150 s, 30 s            \\
\tableline
 Be 70     & U  &  900 s                  & Be 23     & U  & $200 s \times 3$       \\
           & B  & 900 s, 90 s             &           & B  & $200 s \times 3$, 20 s \\
           & V  & 400 s, 40 s             &          & V  & $100 s \times 3$, 10 s  \\
           & I  & 150 s, 12 s             &           & I  & $50 s \times 3$, 5 s   \\
\tableline
 NGC 2259  & U  & $200 s \times 3$        & NGC 2304  & U  &  900 s                 \\
           & B  & $200 s \times 3$, 20 s  &           & B  & 900 s, 90 s            \\
           & V  & $100 s \times 3$, 10 s  &           & V  & 300 s, 30 s            \\
           & I  & $50 s \times 3$, 5 s    &           & I  & 80 s, 7 s              \\
\tableline
\end{tabular}
\end{center}
\end{table*}

\newpage

\pagestyle{empty} 
\begin{table*}
\begin{center}
\tablenum{3}
\caption{Physical parameters of the 12 open clusters.\label{tbl-3}}
\begin{tabular}{ccccc}
\tableline\tableline
Cluster &  $E(B-V)$ & $(m-M)_0$ & [Fe/H] & Age (Myr) \\
\tableline
 Be 104    &   $0.45\pm0.05$   &   $13.2\pm0.2 $   &  ~$~0.07$ dex  &  $~790\pm160$~  \\
 Be 60     &   $0.86\pm0.05$   &   $13.2\pm0.2 $   &  ~$~0.07$ dex  &  $~160\pm60$~~  \\ 
 King 15   &   $0.70\pm0.12$   &   $12.5\pm0.4 $   &  ~$~0.07$ dex  &  $~250\pm50$~~  \\ 
 NGC 381   &   $0.40\pm0.10$   &   $10.3\pm0.3 $   &  ~$~0.07$ dex  &  $~320\pm70$~~  \\ 
 Be 64     &   $1.05\pm0.15$   &   $13.0\pm0.5 $   &   $-0.61$ dex  &  $1000\pm210$~  \\ 
 King 6    &   $0.50\pm0.10$   &   $~9.7\pm0.4 $   &  ~$~0.46$ dex  &  $~250\pm50$~~  \\ 
 NGC 1348  &   $1.02\pm0.05$   &   $11.3\pm0.3 $   &   $~0.07$ dex  &  $~130\pm50$~~  \\ 
 Be 12     &   $0.70\pm0.15$   &   $12.5\pm0.5 $   &   $~0.07$ dex  &  $4000\pm820$~  \\ 
 Be 70     &   $0.48\pm0.10$   &   $13.1\pm0.3 $   &   $-0.32$ dex  &  $4000\pm820$~  \\ 
 Be 23     &   $0.40\pm0.05$   &   $14.2\pm0.3 $   &  ~$~0.07$ dex  &  $~790\pm160$~  \\ 
 NGC 2259  &   $0.59\pm0.05$   &   $12.6\pm0.3 $   &  ~$~0.07$ dex  &  $~320\pm70$~~ \\ 
 NGC 2304  &   $0.10\pm0.02$   &   $13.0\pm0.1 $   &   $-0.32$ dex  &  $~790\pm160$~  \\ 
\tableline
\end{tabular}
\end{center}
\end{table*}

\end{document}